\documentclass[prl,twocolumn,english,showpacs]{revtex4-1}
\usepackage{graphicx}
\usepackage{amssymb}

\newcommand{\ket}[1]{|#1\rangle}

\begin{document}

\title{Decoherence of a single-ion qubit immersed in a spin-polarized atomic bath}

\author{L. Ratschbacher$^{1}$, C. Sias$^{1,2}$, L. Carcagni$^1$, J. M. Silver$^1$, C. Zipkes$^1$, M. K{\"o}hl$^{1,3}$}
\affiliation{$^1$Cavendish\,Laboratory,\,University~of Cambridge, JJ Thomson Avenue, Cambridge CB30HE, United Kingdom\\$^2$Istituto~Nazionale~di~Ottica,~50019~Sesto~Fiorentino,~Italy\\\\$^3$Physikalisches Institut, University of Bonn, Wegelerstrasse 8, 53115 Bonn,~Germany}

\begin{abstract}
We report on the immersion of a spin-qubit encoded in a single trapped ion into a spin-polarized neutral atom environment, which possesses both continuous (motional) and discrete (spin) degrees of freedom. The environment offers the possibility of a precise microscopic description, which allows us to understand dynamics and decoherence from first principles. We observe the spin dynamics of the qubit and measure the decoherence times ($T_1$ and $T_2$), which are determined by the spin-exchange interaction as well as by an unexpectedly strong spin-nonconserving coupling mechanism.
\end{abstract}


\maketitle

A spin-1/2 system represents the most fundamental quantum mechanical object. Its spin-dynamics and -decoherence when interacting with an environment determine its potential use as a qubit and are responsible for a multitude of impurity effects encountered in the solid state. While an extensive amount of theoretical work on this problem exists (for reviews see \cite{Leggett1987,Prokofiev2000}), experiments with well-controlled and adjustable environments are scarce. The necessary sensitivity to probe single spins, ideally with a single-shot readout, is often incompatible with the ability to adjust the properties of the environment. Among the few examples of controlled decoherence processes are the quantum measurement process \cite{Brune1996b}, the decoherence of motional quantum states of trapped ions by noisy classical electric fields \cite{Myatt2000}, and the effects of spontaneous emission \cite{Roos2004,Ozeri2005,Barreiro2011}. These are related to spin-boson physics, where the environment is formed by a set of harmonic oscillator modes \cite{Leggett1987}. On the contrary, the decoherence of a localized spin impurity inside an environment of other spins lies at the heart of the so-called central-spin problem \cite{Prokofiev2000}. Here, the occurring decoherence mechanisms are different from spin-boson physics since the spectrum of the bath is discrete and spin-spin or spin-orbit interactions play a role. The central-spin model has been often applied as a simplified and approximate description of the interaction of semiconductor quantum dots \cite{Hanson2003} and color centres in solids \cite{Hanson2008} with their environment.

Here we investigate the model system of a single localized spin-1/2 coupled to an spin-polarized environment of tunable density. Specifically, we embed a trapped single Yb$^+$ ion, initially laser-cooled to Doppler temperature, into a spin-polarized ultracold neutral bath of $^{87}$Rb atoms and study the resulting decoherence of the ion's internal spin state. We observe an intricate decoherence mechanism, which is spin-nonconserving and involves the coupling of the orbital degrees of freedom with the spin degrees of freedom. We measure the longitudinal ($T_1$) and the transverse ($T_2$) coherence times, also with respect to the energy separation, and identify the time scales for Zeeman- and hyperfine-state relaxation.

\begin{figure}
    \includegraphics[width=1.0\columnwidth]{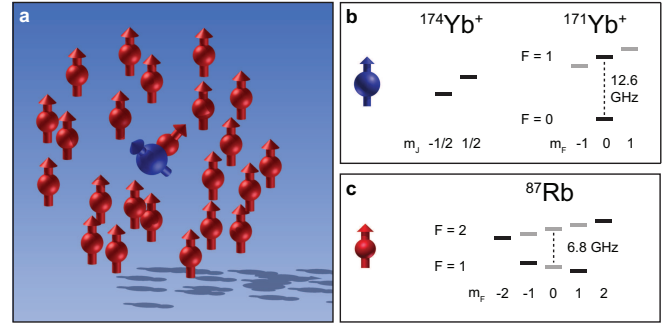}
	\caption[Level Scheme]{\label{fig:levelScheme}(Color online) {\bf a} Illustration of the trapped single ion spin coupled to a spin-polarized neutral atom cloud. {\bf b} Level structure of the Yb$^+$ electronic ground state in a weak magnetic field. To implement the spin-1/2 system, we use either the Zeeman qubit $\ket{m_J=\pm1/2}$ in the isotope $^{174}$Yb$^+$ or the magnetic field insensitive hyperfine qubit $\ket{F=0,m_F=0}$ and $\ket{1,0}$ in the isotope $^{171}$Yb$^+$. {\bf c} The cloud of neutral $^{87}$Rb atoms is prepared in one of the four atomic spin states $\ket{F=2,m_F=2}_a$,$\ket{2,-2}_a$,$\ket{1,1}_a$ or $\ket{1,-1}_a$.}
\end{figure}

The interaction between an ion and a neutral atom is, to leading order, via a central potential $V_\sigma(r)$ where $r$ is the internuclear separation. Asymptotically ($r \rightarrow \infty$), the central potential is determined by the polarization interaction $-C_4/2r^4$ and is spin-independent \cite{Cote2000,Makarov2003,Idziaszek2009}. For small $r$, collisions in the electronic singlet and triplet channels exhibit different potentials $V_S(r)$ and $V_T(r)$ giving rise to a spin-exchange interaction, which is spin-conserving for the total spin of atom and ion. Langevin collisions in the central potential, i.e. collisions with energies above the centrifugal barrier, occur at an energy-independent rate $\gamma_{L} = 2\pi \sqrt{C_4/\mu}\,n_a$, where $\mu$ is the reduced mass and $n_a$ is the neutral atom density. In our experiment we have $\gamma_{L}/n_a =2.1\times10^{-15}$\,m$^3$/s and we use typical densities of $n_a=10^{18}$\,m$^{-3}$. Even though the Langevin rate is significantly smaller than the total collision rate $\gamma_c$ (see Supplementary Material), in previous experiments the Langevin process has dominated cold inelastic ion-atom collisions \cite{Grier2009,Zipkes2010,Zipkes2010b,Schmid2010,Hall2011,Rellergert2011}. For spin-dependent processes additionally the anisotropic magnetic dipole-dipole interaction and the second-order spin-orbit coupling can play important roles, the latter in particular for heavy atoms \cite{Mies1996}. They induce coupling between the spin and the orbital motion and break the conservation of the total electronic spin. In this work we determine which of these mechanisms lead to spin decoherence.

\begin{figure}
    \includegraphics[width=1.0\columnwidth]{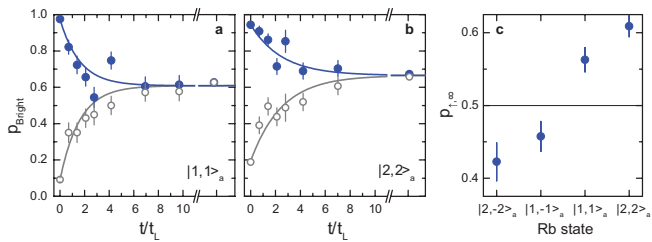}
	\caption[Level Scheme]{\label{fig:figure3}(Color online) Spin-relaxation in the $^{174}$Yb$^+$ Zeeman qubit. The probability $p_{Bright}$ of the ion to occupy the bright $\ket{\uparrow}$ state after preparation in $\ket{\uparrow}$ (full symbols) or $\ket{\downarrow}$ (open symbols) in a bath of {\bf (a)} $\ket{1,1}_a$ or {\bf (b)} $\ket{2,2}_a$  atoms. Error bars denote one standard deviation uncertainty intervals resulting from approximately 3000 measurements per spin state. {\bf c} The equilibrium spin states (corrected for detection efficiencies) for all four atomic bath configurations $\ket{2,-2}_a$, $\ket{1,-1}_a$, $\ket{1,1}_a$ and $\ket{2,2}_a$.}
\end{figure}

First, we study spin relaxation of the Zeeman qubit of the isotope $^{174}$Yb$^+$, which represents an ideal two-level system (see Figure 1). We prepare the $|m_J=1/2\rangle\equiv\ket{\uparrow}$ or the $|m_J=-1/2\rangle\equiv\ket{\downarrow}$ qubit state in an adjustable magnetic field inside the neutral atom cloud. During the interaction period, the ion undergoes binary collisions with the atoms, and we normalize interaction times by the Langevin time constant $t_L=1/\gamma_L$. We detect the population of the qubit state by electronic shelving and determining the probability of the ion being in a bright ($p_{Bright}$) or dark ($p_{Dark}$) state using light scattering on the $S_{1/2}-P_{1/2}$ transition (see Supplementary Material). Measurements of the $T_1$ time are performed for varying density and neutral atom spin composition. The data for the two different bath configurations $\ket{F=1,m_F=1}_a$ and $\ket{2,2}_a$  are shown in Figures 2a and 2b, respectively. We find that the initially polarized ion spin relaxes into a mixed steady-state within a few Langevin collision times. This result cannot be understood from spin-exchange collisions, for which, for example, the doubly spin-polarized combination $\ket{\uparrow}$ and $\ket{2,2}_a$ is protected from spin-changing collisions due to spin conservation. From our steady-state population data for atoms in the $\ket{2,2}_a$ state and the measured $T_1 = (2.50 \pm 0.39)t_L$, we determine the spin-exchange and spin-relaxation rates to be $\gamma_{\downarrow,SE} = (0.22\pm0.03) /T_1$ and $\gamma_{\uparrow,SR} = \gamma_{\downarrow,SR} = (0.39\pm0.02)/T_1$ (see Supplementary Material). An analogous result holds for the $\ket{2,-2}_a$ state. Notably, the measured spin-relaxing collision rate is approximately 5 orders of magnitude higher than the charge-exchange collision rate \cite{Zipkes2010b,Ratschbacher2012}. All measurements were performed at an energy splitting of the ion Zeeman qubit of 37.5\,MHz, however the same behaviour was found for different Zeeman splittings in the range from 0.39\,MHz to 139\,MHz in a $\ket{2,2}_a$ environment.

The observed spin dynamics of the Yb$^+$ ion in a spin polarized cloud of $^{87}$Rb atoms is inconsistent with the picture of dominant spin-exchange and negligible spin-relaxation, that has previously been reported in He$^+$ + Cs collisions \cite{Major1968}, in neutral atom-atom collisions even at room temperature \cite{Baranga1998}, and in semiconductor quantum dots \cite{Latta2008}. Moreover, investigations in both ultracold gases \cite{Soding1998} and optically pumped vapor cells \cite{Kadlecek2001} have revealed that even for heavy alkali-metal collisions the hierarchy $\gamma_c \approx \gamma_{SE} \gg \gamma_{SR}$ is fulfilled. In such a situation, the dominant spin-exchange interaction leads to a steady state near a perfect polarization of spin and environment both when the environment is initially polarized but the spin is unpolarized and when the single spin is continuously pumped but the environment is initially unpolarized. The strong spin-relaxation observed here could be aided by a level crossing of both the incoming singlet $A^1\Sigma^+$ and triplet $a^3\Sigma^+$ channels to a $^3\Pi$ channel at short internuclear separation \cite{Dulieu2012,Buchachenko2012}, which potentially provides a  mechanism for spin-orbit coupling. A conceptually similar competition between spin-orbit induced relaxation and spin-conserving processes seems to play a role in the ultrafast relaxation of magnetization in ferromagnets \cite{Koopmans2010}.

The dependence of the steady-state spin population of the ion on the atomic bath configuration has been determined for the $\ket{2,-2}_a$, $\ket{1,-1}_a$, $\ket{1,1}_a$, and $\ket{2,2}_a$ states and is shown in Figure 2c. The steady-state values for the $\ket{1,-1}_a$ and $\ket{1,1}_a$ states can be qualitatively understood considering their projections into the electron singlet and triplet bases and taking into account that spin-exchange collisions leading to a transition of the atomic hyperfine state from the $\ket{F=1}_a$ to the $\ket{F=2}_a$ manifold are energetically suppressed.

\begin{figure}
    \includegraphics[width=0.95\columnwidth]{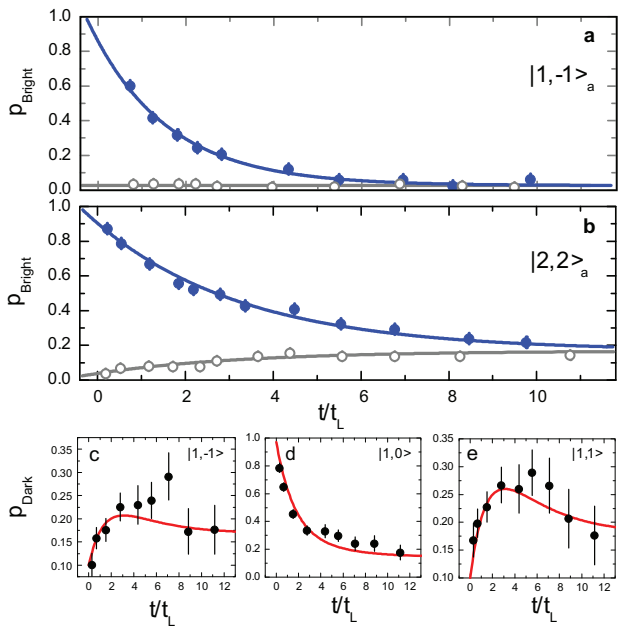}
	\caption[Level Scheme]{\label{fig:figure2}(Color online) Hyperfine spin relaxation in $^{171}$Yb$^+$. {\bf a} The probability $p_{Bright}$ after preparation in the $\ket{1,0}$ (full symbols) or the $\ket{0,0}$ (open symbols) state vs. the interaction time for atoms in the $\ket{1,-1}_a$ state. Error bars denote one standard deviation uncertainty intervals resulting from a total of 5000 measurements. The fit values at $t=0$ are limited by detection errors. {\bf b} Similar data for collisions with $^{87}$Rb atoms in the $\ket{2,2}_a$ state and a total of 19000 measurements. {\bf c-e} Zeeman-resolved detection within the $F=1$ manifold after preparation in $\ket{1,0}$ (atoms in $\ket{2,2}_a$) with 1200 measurements per Zeeman state. The measurements are performed by applying resonant $\pi$ pulses, exchanging the population of the dark state $\ket{0,0}$ with $\ket{1,-1}$, $\ket{1,0}$ or $\ket{1,1}$ immediately before the detection of the probability of the ion being in the dark state $p_{Dark}$.}
\end{figure}

Next, we study the spin relaxation of the magnetic field insensitive hyperfine qubit of the electronic ground state of $^{171}$Yb$^+$ (see Figure 1 and Supplementary Material). The spin detection is performed by resonant light scattering on the $S_{1/2}\ket{F=1}\rightarrow\,P_{1/2}\ket{F=0}$ transition and measuring the probability $p_{Bright}$ ($p_{Dark}$) of the ion being in a bright (dark) state $\ket{F=1}$  ($\ket{F=0}$). If the neutral atoms are prepared in the $\ket{1,-1}_a$ state (see Figure 3a), the ion relaxes towards the spin ground state, as expected for amplitude decoherence, and we observe an exponential decay with a time constant $T_1$ of a few Langevin collision times.

In contrast, if the neutral atoms are prepared in the $\ket{2,2}_a$ state (Figure 3b), the steady state probability to measure the ion spin in the $\ket{F=1}$ manifold is 0.16(1). We interpret this non-zero steady-state value as resulting from the intake of hyperfine energy from the atoms. In order to quantitatively assess this effect, we assign the steady-state distribution of the ion's hyperfine spin state a ``spin-temperature'' $T_s$ by $\frac{p_1}{p_0}=3\exp\left(-\frac{E^{HFS}}{k_B T_s}\right)$ resulting in $T_s=200$\,mK. Here $p_1=\sum_m p_{\ket{1,m}}$ is the probability $p_{Bright}$ corrected for detection efficiencies (see Supplementary Material) and $E^{HFS}=h\times 12.6$\,GHz is the hyperfine energy of the ion. In comparison, we estimate the kinetic energy $E_{kin}$ from the balance of elastic and inelastic processes, assuming to be not dominated by micromotion heating effects at these temperatures \cite{Zipkes2011}. The average kinetic energy intake of the ion per Langevin collision due to a hyperfine flip in the neutral atom is $\delta E_{heat}=\epsilon E_a^{HFS} \frac{m_a}{m_a+m_i}$. Here, $m_a$ is the mass of the atom, $m_i$ the mass of the ion, $0\leq \epsilon \leq 1$ describes the probability that an atomic hyperfine flip occurs during a Langevin collision, and $E_a^{HFS}=h\times6.8$\,GHz is the internal hyperfine energy of the atom. The average ion energy loss by elastic ion-atom momentum transfer is $\delta E_{cool} = -\frac{2m_a m_i}{(m_a+m_i)^2}E_{kin}$ per Langevin collision. As a result, the steady-state average kinetic energy of the ion is $ \langle E_{kin} \rangle = \epsilon E_a^{HFS} \frac{m_a+m_i}{2m_i}=\epsilon k_B\times240$\,mK. For $\epsilon$ near unity, we hence obtain $\langle E_{kin} \rangle \approx k_B T_s$, which signals an equidistribution of energy between kinetic and spin degrees of freedom. In contrast, if the atomic bath is prepared in its hyperfine ground state $\ket{F=1}_a$, no energy-releasing hyperfine changes can occur and the temperature of ion is expected to be limited by micromotion heating, which is on the order of 20\,mK \cite{Major1968,Zipkes2011}.

As an additional degree of freedom in the case of $^{171}$Yb$^+$, we study the spin transfer within the $\ket{F=1}$ manifold of $^{171}$Yb$^+$ when starting from $\ket{1,0}$. The data are shown in Figure 3c-e, together with the results of a four-level rate equation model, which involves spin transfer within the  $\ket{F=1}$ manifold as well as decay into the $\ket{F=0}$ state. Most strikingly, the decay out of the initially prepared $\ket{1,0}$ state exhibits two different time constants (see Figure 3d). The fast initial decay is associated with the population of the $\ket{1,\pm1}$ states, where the occupation correspondingly rises (see Figure 3c and e). Eventually, the $\ket{F=1}$ states decay to their steady-state populations. The significant build-up of population in the $\ket{1,-1}$ state and the relaxation from $\ket{1,1} \rightarrow \ket{0,0}$ are both forbidden by spin conservation, however, so as for the Zeeman qubit we observe a strong spin-nonconserving process.

Quantum mechanically, an isolated spin-1/2 can exist in superposition states of spin-up and spin-down that are not describable by classical physics. Coupling to an environment affects the quantum mechanical superposition and, eventually, leads to decoherence as the information of the quantum correlations gets lost and the quantum mechanical superpositions transform into probability distributions of occupation numbers. In the problem discussed here, it is important to understand whether this decoherence results from inelastic spin-relaxation processes, i.e. amplitude decoherence, or whether also elastic (forward) scattering contributes to additional phase decoherence of superposition states. In the hydrogen maser \cite{Koelman1988} spin-exchange processes have been identified as a leading source of both line shifts and broadening.

We characterize the influence of ion-atom collisions on the hyperfine clock transition in $^{171}$Yb$^+$ by measuring its coherence time and line shift using the Ramsey technique. To this end, we prepare the ion in the $\ket{0,0}$ hyperfine ground state and apply a $\pi/2$-pulse on the 12.6-GHz clock transition, creating a superposition state $(\ket{0,0}+i\ket{1,0})/\sqrt{2}$. After a waiting time of 27\,ms we apply a second $\pi/2$-pulse, followed by a readout of the $\ket{F=1}$ spin state. Figure 4 displays our results. In panels a-d we show sample traces of our Ramsey fringes as a function of the detuning of the microwave pulse from the clock transition with increasing density of neutral atoms during the interaction time. Even without neutral atoms, the visibility of the Ramsey fringes is limited to $(55\pm2)\%$ owing to magnetic field fluctuations of the bias field. We find that the spin coherence decays within $T_2=(1.4\pm0.2)t_L$ (see Figure 4e). This is on the time scale of the population relaxation of the $\ket{1,0}$ state, which is driven by the spin-exchange and spin-orbit interaction. The measurement hence identifies spin-relaxation as the leading mechanism of spin decoherence in our system with a non-detectable contribution from elastic (forward) scattering. Finally, we note that the frequency shift $\Delta \nu$ of the clock transition is below the resolution set by the spin relaxation rate and is $\Delta \nu \leq 4\times 10^{-11} E_i^{HFS}/h$ for our densities (see Figure 4f).

Our measurements realize a novel spin-bath with additional continuous degrees of freedom in which we have observed an unexpectedly strong coupling between internal (spin) and external (orbital) degrees of freedom. Future experiments with a driven impurity could reveal whether this coupling mechanism could lead to similar quantum phase transitions as predicted for the spin-boson model \cite{Leggett1987} and how the physics of the single ion in the spin-polarized bath relates to the physics of polarons \cite{Schirotzek2009,Mora2010,Kohstall2012,Koschorreck2012}.

We thank A. Imamoglu and C. Kollath for discussions. The work has been supported by {EPSRC} (EP/H005676/1), ERC (grant 240335), the Leverhulme Trust (CS), the Royal Society, and the Wolfson Foundation.

\begin{figure}
    \includegraphics[width=1.0\columnwidth]{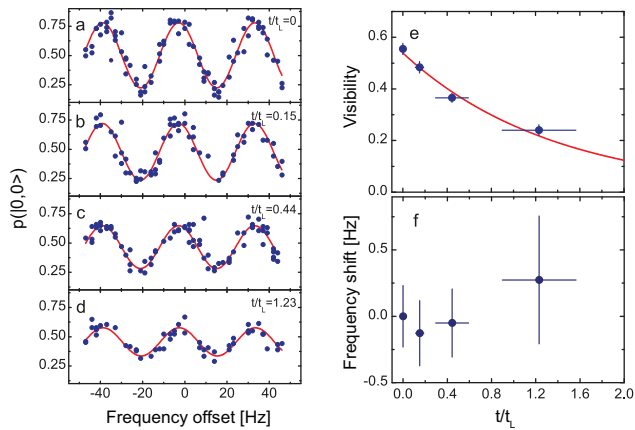}
	\caption[Coherence]{\label{fig:figure4}(Color online)  Spin coherence of the hyperfine clock transition $\ket{0,0} \leftrightarrow \ket{1,0}$ in $^{171}$Yb$^+$. \textbf{a}-\textbf{d} Ramsey fringes recorded for increasing atomic density. \textbf{e} Ramsey contrast as a function of $t/t_L$ displays an exponential decay with a time constant of $T_2=(1.4\pm0.2)t_L$. \textbf{f} No frequency shift of the clock transition is observed within the experimental errors. Error bars denote one standard deviation uncertainty intervals.}
\end{figure}

\newpage

\section{Supplementary material}

\subsection{Experimental setup}
We prepare up to $8 \times 10^5$ neutral $^{87}$Rb atoms in the $|F=2,m_F=2\rangle_a$ hyperfine state of the electronic ground state at temperatures down to $T\approx 200$\,nK in a harmonic magnetic trap of characteristic frequencies \cite{Palzer2009} $(\omega_x,\omega_y,\omega_z)=2 \pi \times (8,26,27)$\,Hz. When prepared in the $\ket{F=1, m_F=-1}_a$ state the trap frequencies are $(\omega_x,\omega_y,\omega_z)=2 \pi \times (5,10,15)$\,Hz. Alternatively, the atoms can be transferred into an optical dipole trap formed by two crossed laser beams at 1064\,nm. Here, the atoms can also be trapped in the hyperfine ground state $\ket{1,1}_a$ or the $\ket{2,-2}_a$ state. At the same location, we trap a single Yb$^+$ ion in a radio-frequency Paul trap with secular trap frequencies of $\omega_\perp=2 \pi \times 150\,$kHz radially and $\omega_{ax}=2 \pi \times 55$\,kHz axially \cite{Zipkes2010,Zipkes2010b}. The atomic density at the location of the ion in the centre of the neutral bath is determined for magnetically trapped clouds from  atom number and temperature measured in absorption imaging and the harmonic trap frequencies. For optically trapped clouds, the atomic density is determined by measuring the inelastic loss rate of the ion in its electronically excited $D_{3/2}$-state \cite{Ratschbacher2012}. In both cases we estimate the uncertainty of the absolute atomic densities to be about 40\%. In order to avoid losses of the ion due to inelastic collisions in the highly reactive electronically excited states \cite{Ratschbacher2012}, the atoms are removed from the position of the ion before detection of the $^{174}$Yb$^+$ qubit. For steady-state spin measurements the same atomic cloud has been used for up to eight consecutive measurements by displacing and returning the atoms by moving the position of the optical trap beams.

\subsection{Ion-atom interaction}
The interaction between an ion and a neutral atom at long distances is dominated by the attractive polarization interaction potential, which is of the form of a central potential $V(r) = -\frac{C_4}{2 r^4}$. Here, $C_4=\alpha_0 q^2/(4 \pi \epsilon_0)^2$ is proportional to the neutral particle polarizability $\alpha_0$, $q$ is the charge of the ion, $\epsilon_0$ is the vacuum permittivity, and $r$ is the internuclear separation. The total collision rate between atoms and ions in the central potential is given by $\gamma_c =n_a \sqrt{2} \pi(1+\pi^2/16) (C_4/\hbar)^{2/3}  (E/\mu)^{1/6}$ and includes quantum mechanical forward-scattering \cite{Cote2000}. Here, $E$ is the collision energy. In our experiment we have $\gamma_{L}/n_a =2.1\times10^{-15}$\,m$^3$/s and $\gamma_c/n_a=2.5\times10^{-14}$\,m$^3$/s for a collision energy of $E=k_B \times 100$\,mK.

\subsection{Ion qubit preparation and detection}

Ultraviolet light at 370\,nm, resonant with $S_{1/2}\rightarrow\,P_{1/2}$ transition, is used for laser cooling and spin state preparation, while laser light at 935\,nm clears out the  $^2D_{3/2}$ state. State preparation in either Zeeman state of the $S_{1/2}$ ground state of the $^{174}$Yb$^{+}$ isotope is achieved by optical pumping with $\sigma^+-$ or $\sigma^--$ polarized light on the $S_{1/2} \rightarrow\,P_{1/2}$ transition. State detection is performed by frequency-selective electron shelving in the $F_{7/2}$ state via the $D_{5/2}$ state using a laser at 411\,nm and subsequent resonant light scattering on the $S_{1/2} \rightarrow\,P_{1/2}$ transition \cite{Ratschbacher2012}. Imperfections in the detection scheme limit the direct correspondence of the detection of a bright or dark quantum state, as determined from the statistics of photon events, and the occupation of a specific upper or lower spin quantum state. The probability of spin occupation $p_{\downarrow}$ is determined by $p_{\downarrow} = (p_{Dark}-\eta_{Dark,\uparrow})/(\eta_{Dark,\downarrow}-\eta_{Dark,\uparrow})$ from the detected distribution of dark states $p_{Dark}$ and the independently determined detection errors $(1-\eta_{Dark,\downarrow})$ and $\eta_{Dark,\uparrow}$. For measurements with $^{87}$Rb $\ket{F=2}_a$, the detection errors are $(1-\eta_{Dark,\downarrow}) = 0.19\pm 0.01$ and $\eta_{Dark,\uparrow} = 0.03 \pm0.01$, and for measurements involving the $\ket{F=1}_a$ manifold we find $(1-\eta_{Dark,\downarrow}) = 0.10 \pm 0.01$ and $\eta_{Dark,\uparrow} = 0.00 +0.01$. The larger detection errors of the Zeeman qubit when the atoms are in the $\ket{F=2}_a$ state result from the increased ion kinetic energy due to release of atomic hyperfine energy.

For the isotope $^{171}$Yb$^{+}$, state preparation in the $\ket{0,0}$ state is achieved by a short (250\,$\mu$s) laser pulse resonant with the $S_{1/2} \ket{F=1}\rightarrow \,P_{1/2} \ket{F= 1}$ transition \cite{Olmschenk2007}. Microwave radiation resonant with the $12.6$-GHz hyperfine transitions is used for coherent qubit manipulation on the clock transition, for Rabi $\pi$-flips on the $\ket{0,0}\rightarrow\ket{1,-1}$ and $\ket{0,0} \rightarrow\ket{1,1}$ transitions as well as for repumping during Doppler cooling. The magnetic bias field strength depends on the atom trap configuration and ranges from $5.64$\,G in the optical trap to $7.56$\,G in the magnetic trap of the $^{87}$Rb$\ket{1,-1}_a$ atoms. The spin detection probes fluorescence of the ion on the $S_{1/2}\, \ket{F=1} \rightarrow P_{1/2}\, \ket{F=0}$ transition and we use a standard statistical analysis of photon detection events for state discrimination \cite{Acton2006}. The fluorescence detection can distinguish the states $\ket{0,0}$ and $\ket{F=1}$ and the detection errors are $(1-\eta_{Dark,\downarrow}) = 0.02 \pm 0.01 $ for the hyperfine ground state $\ket{F=0}$ and $\eta_{Dark,\uparrow} = 0.07 \pm 0.03$ for the $\ket{F=1}$ hyperfine states.

\subsection{Determination of spin-exchange and spin-relaxation rates}
The spin relaxation behaviour is well described by solutions of the general two-level rate equation model $p_{\uparrow}(t) = p_{\uparrow,0} e^{-t/T_1} + p_{\uparrow,\infty} (1-e^{-t/T_1})$ with the longitudinal coherence time $T_1=1/(\gamma_{\uparrow}+\gamma_{\downarrow})$ and the equilibrium spin state $ p_{\uparrow, \infty} =  \gamma_{\uparrow}/(\gamma_{\uparrow}+\gamma_{\downarrow})$. The decay rates for the spin states $\gamma_{\uparrow}=\gamma_{\uparrow,SE}+\gamma_{\uparrow,SR}$ and $\gamma_{\downarrow}=\gamma_{\downarrow,SE}+\gamma_{\downarrow,SR}$ are decomposed into a spin-exchange part depending on the atomic bath spin and a spin-relaxation part including spin-orbit coupling. For collision with the maximally polarized bath state $\ket{2,2}_a$ at energies much larger than the Zeeman splittings, the relations $\gamma_{\uparrow,SE} = 0$ and $\gamma_{\downarrow,SR}=\gamma_{\uparrow,SR}$ hold true and result in $\gamma_{\downarrow,SE} =  (2p_{\uparrow, \infty}-1)/T_1 = (0.22\pm0.03)/T_1$ and $\gamma_{\uparrow,SR} = \gamma_{\downarrow,SR} = (1-p_{\uparrow, \infty})/T_1 = (0.39\pm0.02)/T_1$.

\begin{table*}[ht]
\begin{center}
\begin{tabular}{|l|l|l|l|l|l|l|l|}
\hline
 	     Ion isotope        &  Atom state & $T_1/t_L$ &$p_{\uparrow,\infty}$ & $\sum p_{\ket{1,x},\infty}$ & $T_2/t_L$ \\
 	\hline
 	$^{171}$Yb$^{+}$   &$\ket{1,-1}_a$ &$1.73\pm 0.17$ &  & $0.000 + 	0.005$  &\\
 	$^{171}$Yb$^{+}$   &$\ket{2,2}_a $ &$3.39\pm 0.16$ &  & $0.163 \pm 	0.013$ &$1.4\pm0.2 $\\
 	$^{174}$Yb$^{+}$   &$\ket{2,2}_a$  &$2.50\pm 0.39$ & $0.609 \pm 0.015$ &  &\\
 	$^{174}$Yb$^{+}$   &$\ket{2,-2}_a$ & & $0.423 \pm 	0.026$ &  &\\
    $^{174}$Yb$^{+}$   &$\ket{1,1}_a$  &$1.60\pm 0.24$ & $0.563 \pm 0.017$ & &\\
    $^{174}$Yb$^{+}$   &$\ket{1,-1}_a$ & & $0.457 \pm 	0.021$ &  &\\
 	\hline
\end{tabular}
\caption{Summary of the decoherence times ($T_1$ and $T_2$) and steady-state spin distributions of all spin-bath combinations. In addition to the statistical errors quoted, all times have systematic uncertainties of $40\%$ due to the determination of the absolute atomic densities at the location of the ion. All values are compensated for detection efficiencies.}
\end{center}
\end{table*}

\end{document}